\documentclass[aps,prd,twocolumn,groupedaddress,showpacs]{revtex4}

\usepackage{graphicx}

\begin{document}

%\preprint{}

\title{Consequences of statistical sense determination 
for WIMP directional detection}


\author{Anne M. Green}
\email[]{anne.green@nottingham.ac.uk}
%\thanks{}
\affiliation{School of Physics and Astronomy, University of 
  Nottingham, University Park, Nottingham, NG7 2RD, UK}
\author{Ben Morgan}
\email[]{ben.morgan@warwick.ac.uk}
%\thanks{}
\affiliation{Department of Physics, University of Warwick, Coventry, 
CV4 7AL, UK}


\date{\today}

\begin{abstract}
  We study the consequences of limited recoil sense reconstruction on the
  number of events required to reject isotropy and detect a WIMP
  signal using a directional detector. For a constant probability of
  determining the sense correctly, 3-d read-out and zero background,
  we find that as the probability is decreased from 1.0 to 0.75 the
  number of events required to reject isotropy using the mean angle
  statistic is increased by a factor of a few. As the probability is
  decreased further the number of events required using this statistic
  increases sharply, and in fact isotropy can be rejected more easily
  by discarding the sense information and using axial statistics. This
  however requires an order of magnitude more events than vectorial
  data with perfect sense determination. We also consider energy
  dependent probabilities of correctly measuring the sense, 2-d
  read-out and non-zero background. Our main conclusion regarding the
  sense determination is that correctly determining the sense of the
  abundant, but less anisotropic, low energy recoils is most important
  for minimising the number of events required.

\end{abstract}

\pacs{95.35.+d}
%\keywords{}

\maketitle

%\section{}
\section{Introduction}
%\subsection{}
%\subsubsection{}

Weakly Interacting Massive Particle (WIMP) direct detection
experiments aim to directly detect non-baryonic dark matter via the
elastic scattering of WIMPs on detector nuclei~\cite{DD}, and are
presently reaching the sensitivity required to detect the lightest
neutralino (which in most supersymmetry models is the lightest
supersymmetric particle and an excellent WIMP candidate). Since the
expected event rates are very small ( ${\cal O} (10^{-5} - 1)$ counts
${\rm kg^{-1} day^{-1}}$) distinguishing a putative WIMP signal from
backgrounds due to, for instance, neutrons from cosmic-ray induced
muons or natural radioactivity, is crucial. The direction dependence
of the WIMP scattering rate (due to the Earth's motion with respect to
the Galactic rest frame)~\cite{dirndep} provides a potential WIMP
`smoking gun' and low pressure gas time projection chambers (TPCs),
such as DRIFT ({\bf D}irectional {\bf R}ecoil {\bf I}dentification
{\bf F}rom {\bf T}racks)~\cite{drift,sean:drift} and
NEWAGE~\cite{newage}, see also Ref.~\cite{martoff}, seem to offer the
best prospects for a detector capable of measuring the directions of
sub-100 keV nuclear recoils.

Early studies found that in principle as few as 30 events would be
required to distinguish a WIMP induced signal from isotropic
backgrounds~\cite{copi:krauss,lehner:dir}. In reality the number of
events, and hence the exposure required, depends on the detector
properties including the energy threshold, background event rate,
whether the read-out measures the recoil momentum in 2 or 3 dimensions
(and if 2-d in which plane)~\cite{pap1,pap2,copi2d,pap3} and whether
the sense (i.e. the absolute sign $+ {\vec{p}}$ or $- { \vec{p}}$) of
the recoil momentum vectors can be
measured~\cite{pap1,pap2,copi2d,pap3,vf}.

The factor
with the biggest effect on the number of events required is whether or
not the sense of the recoils can be measured~\cite{pap1,pap2,pap3}. 
If the data is axial (i.e. the sense can not be measured) the exposure
is increased, compared to the vectorial case, by one order of magnitude for
3-d read-out and at least two orders of magnitude for 2-d read-out in the
optimal plane~\footnote{This can be improved to a factor $\sim 30$ if the
  reduced angles (with the direction of solar motion subtracted) are
  calculated and analyzed.}~\cite{pap3}. At low energies, the energy deposition
dE/dx of nuclear recoils, and hence the density of ionisation created, is
predicted to slowly decrease with decreasing recoil
energy~\cite{hitachi-dedx}. Thus the sense of a recoil track is, in principle,
measurable by determining the direction in which the ionisation density
decreases along the track. 

Previous work on simulating recoil tracks using SRIM2003~\cite{srim} indicated
that the ionization density distributions reconstructed from recoil tracks
were close to uniform due to both fluctuations in the production of
ionisation and diffusion during the drift of this ionisation to the
read-out plane~\cite{pap1,pap3}.  It was therefore not possible to
determine the absolute signs of the reconstructed recoil vectors in these
simulations. It should be noted however that while SRIM2003 predicts
sulfur recoil ranges and quenching factors (fraction of recoil energy
going into ionisation) in agreement with experimental
data~\cite{drift:neutron}, it was not designed to model recoils in
gaseous targets.

Whether or not the sense of low energy nuclear recoils can be
determined is therefore an important question which needs to be
resolved experimentally. Ongoing studies~\cite{cygnus,senseexpt}
indicate that the expected decrease in dE/dx is observable, but
fluctuations in the ionization distribution caused by the ionization
process and diffusion may wash out the information in the ionization
distribution leading to errors in sense reconstruction. In other
words, if the sense of a recoil of energy $E$ is determined via a
hypothesis test on some parameter $M$ derived from its reconstructed
ionization distribution, then the PDFs $g(M| {\rm forward \,
  sense},E)$ and $g(M| {\rm backward \, sense},E)$ will not be cleanly
separated, i.e. the significance level, $\mathcal{S}(E)$, is greater
than zero and the power, $\mathcal{P}(E)$, is less than one. Here, the
significance level is the probability of rejecting the forward sense
hypothesis when it is true, and the power is the probability of
rejecting the backward sense hypothesis when it is
false~\cite{cowan:stats}. Provided $\mathcal{S}(E)<0.5$ and
$\mathcal{P}(E)>0.5$, the correct recoil sense will be reconstructed
more often than not with probability $P({\rm correct \, sense}
|E)=1-\mathcal{S}(E)$. In this Brief Report we therefore study the
effect of this probabilistic sense reconstruction on the number of
events required to distinguish a WIMP induced recoil signal from
isotropic backgrounds.

\section{Calculations}
We use the same statistical techniques and methods for calculating the
directional nuclear recoil spectrum as in Refs.~\cite{pap1,pap2,pap3}.
We briefly summarise these procedures here, for further details see
these references and Ref.~\cite{bm:thesis}. We consider the simplest
possible model for the Milky Way halo, an isotropic sphere with local
density $\rho=0.3 \, {\rm GeV} \, {\rm cm}^{-3}$ and a Maxwellian
velocity distribution with dispersion $\sigma_{\rm v}=270 \, {\rm km
  \, s}^{-1}$, and fix the WIMP mass at $m_{\chi}=100 \, {\rm keV}$.
Our simulated detector is a TPC filled with 0.05 bar CS$_2$ gas, a
200 $\mu$m pitch micropixel readout plane, a 10 cm drift length over
which a uniform electric field of 1 kV cm$^{-1}$ is applied, and is
based on the design of the DRIFT-I/II detector~\cite{sean:drift}.  We
use the SRIM2003~\cite{srim} package to generate sulfur recoil tracks
and for 3-d read-out recoil directions are reconstructed as the
principal axis of the charge distributions recorded by the pixels.
2-d read-out would measure the projection of the recoil momentum
vector into a plane fixed on the Earth. The degree of anisotropy of
the 2-d recoil angles (and hence the detectability of a WIMP signal)
depends on the orientation of the read-out
plane~\cite{pap2,copi2d,pap3}. Here we focus on the optimal case of a
read-out plane with normal perpendicular to the spin axis of the
Earth. No simulations of the angular resolution function of a 2-d
detector are available~\footnote{A 2-d read-out projects the recoil
  track into a plane and the effects of this combined with multiple
  scattering and diffusion will make the angular resolution a function
  of both the energy and primary recoil direction.} we therefore
assume perfect resolution in this case (and hence the resulting
numbers of events are lower limits on the number which would be
required in reality). For primary recoil energies below $E_{\rm th}=20
\, {\rm keV}$ the short track length (3-4 pixels) and multiple
scattering make it impossible to reconstruct the track direction in
our simulated detector, we therefore only consider events with
energies above this threshold. Zero background is the goal of the next
generation of experiments made from low activity materials with
efficient gamma rejection and shielding, located deep underground
~\cite{drift2:design}, however we investigate the effect of non-zero
isotropic background by varying the ratio of the background and signal
event rates.

Recoil directions in 3-d and 2-d constitute points on the unit sphere
and circle respectively. For 3-d data the most powerful test for
rejecting isotropy uses the average of the cosine of the angle between
the direction of solar motion and the recoil direction, $\langle
\cos{\theta} \rangle$~\cite{pap1}.  For 2-d data the most powerful
test is the Rayleigh test~\cite{rayleigh} which uses the mean
resultant length of the projected recoil vectors which, modulo
fluctuations, should be zero for data drawn from an isotropic
distribution~\cite{pap2}. 

We calculate the probability distribution of
the relevant statistic, for a given number of events $N$, by Monte
Carlo generating $10^5$ experiments each observing $N$ recoils drawn
from our simulated 3-d or 2-d distributions.  We then compare this
with the null distribution of the statistic, under the assumption of
isotropy and calculate the rejection and acceptance factors, $R$ and
$A$, at each value $T$ of the statistic. The rejection factor is the
probability of measuring a value of the statistic less than $T$ if the
null (isotropic) hypothesis is true or equivalently the confidence
with which the null hypothesis can be rejected given that measured
value of the statistic.  The acceptance is the probability of
measuring a value of the statistic larger than $T$ if the alternative
hypothesis is true or equivalently the fraction of experiments in
which the alternative hypothesis is true that measure a larger
absolute value of the test statistic and hence reject the null
hypothesis at confidence level $R$. Clearly a high value of $R$ is
required to reject the null hypothesis, while a high $A$ is also required,
otherwise any one experiment might not be able to reject the null
hypothesis at the given $R$ or the null hypothesis might be
erroneously rejected. We therefore find the number of events required
for $A_{\rm c}=R_{\rm c}=0.95$, $N_{95}$.

We consider several functional forms for the energy dependence of the
probability of correctly determining the recoil sense, $P({\rm correct
  \, sense}|E)$ (hereafter $P_{cs}(E)$). Our simplest model assumes an
energy independent probability, $P_{cs}(E)=P_{100 \, {\rm keV}}$, with
$0.55 \leq P_{100 \, {\rm keV}} \leq 1.0$. In reality it is likely that
it will be easier to determine the sense of higher energy recoils due
to their longer track lengths (c.f. Ref.~\cite{senseexpt}). We
therefore also consider linearly increasing probability
$P_{cs}(E)=aE+b$. We parametrise this function in terms of the values
of $P_{cs}(E_{\rm th}=20 \, {\rm keV})$, where $E_{\rm th}$ is the
threshold energy as above, and $P_{cs}(E=100 \, {\rm keV})=P_{100 \, {\rm
    keV}}$.  We consider two parameter sets for this parametrization,
$P_{cs}(E_{\rm th})=0.50$ with $0.55\leq P_{100 \, {\rm keV}} \leq
1.0$, and $P_{cs}(E_{\rm th})=0.75$ with $0.75\leq P_{100 \, {\rm
    keV}} \leq 1.0$, the latter being more optimistic about
reconstruction at low energies. In each case for
$E>E_{\rm lim}$, where $P_{cs}(E_{\rm lim})=1.0$, we set
$P_{cs}(E>E_{\rm lim})=1.0$.

 \begin{figure}
 \includegraphics[width=8.0cm]{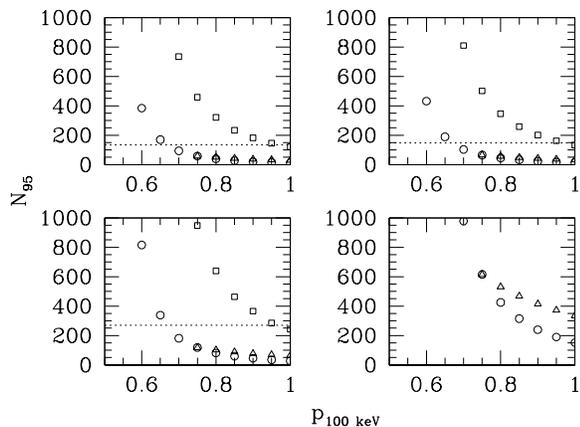}
 \caption{\label{all}The dependence of the number of WIMP events required
   to reject isotropy at 95\% confidence in 95\% of experiments,
   $N_{95}$, using the value of $\langle \cos{\theta} \rangle$ on the
   probability of correctly determining the recoil sense. 
 The circles are for energy independent
   probability, $P_{cs}(E)=P_{100 {\rm keV}}$, triangles (squares) for the
   probability linearly increasing from $P_{cs}(E_{\rm th}=20 \, {\rm keV})=0.75
 \,  (0.5)$ to $P_{cs}(E=100 \, {\rm keV})$= $P_{100 {\rm keV}}$
  The dotted line
  shows the number of events required using the value of $\langle |\cos{\theta}|
 \rangle$, which does not require the sense of the recoils. 
The panels are (top row from left to right and then bottom row) zero background
and signal to (isotropic) background ratio $S/N=10, \, 1$ and $0.1$.
 In all cases, including the axial $\langle |\cos{\theta}|
 \rangle$ statistic,
 we assume 3-d readout with an energy threshold of
 $E_{\rm th}=20 \, {\rm keV}$ and take into account the uncertainty in the
 reconstruction of the recoil direction.}
 \end{figure}

 \begin{figure}
 \includegraphics[width=8.0cm]{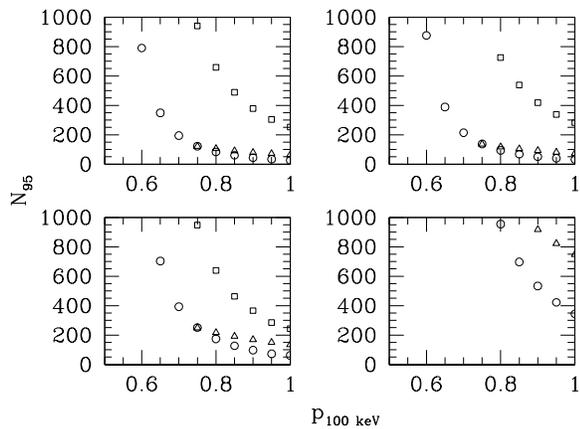}
 \caption{\label{all2d} As fig.~\ref{all} for 2-d read-out using the Rayleigh
statistic. As discussed in the text the angular resolution is not taken into
account in this case, and these numbers are hence lower limits on the
number of events required by a real detector.}
 \end{figure}

\section{Results}

For 3-d read-out, zero background and constant sense determination
probability the number of events required to reject isotropy using the
$\langle \cos{\theta} \rangle$ statistic initially decreases fairly
slowly as the probability is decreased (from $N_{95}=11$ for $P_{100
  \, {\rm keV}}=1.0$ to $N_{95}=39$ for $P_{100 \, {\rm keV}}=0.8$).
The increase becomes more rapid as $P_{100 \, {\rm keV}}$ is decreased
further and $N_{95} \rightarrow \infty$ as $P_{100 \, {\rm keV}}
\rightarrow 0.5$. The linearly increasing sense determination
probability with $P_{cs}(E_{\rm th})=0.75$ and $P_{100 \, {\rm keV}}=
1.0 \, (0.75)$ requires $\sim 3$ times more (the same number of)
events as the constant probability with the same value of $P_{100 \,
  {\rm keV}}$. For $P_{cs}(E_{\rm th})=0.50$ the increase is more
dramatic (more than an order of magnitude for $P_{100 \, {\rm keV}}=
1.0$).  In fact in this case (and also for the constant probability,
if $P_{100 \, {\rm keV}}< \sim 0.7$) fewer events are needed to reject
isotropy using the axial $\langle |\cos{\theta}| \rangle$ statistic
which does not use any information about the sense of the recoil. 
  This is at first glance surprising, however in these cases the low
  values of $P_{cs}$, especially at low energies where the recoil rate
  is higher and the anisotropy in directions lower, significantly
  decrease the anisotropy in the observed recoil momentum vectors and
  hence the sensitivity of the $\langle \cos{\theta} \rangle$
  statistic. This does not affect the $\langle |\cos{\theta}| \rangle$
  statistic, as it is independent of the value of $P_{cs}$, being
  sensitive to the concentration of recoil axes, rather than vectors,
  around the line of solar motion. The number of events required using
the axial statistic is,
however, an order of magnitude larger than for vectorial data with
perfect sense determination. As the signal to noise ratio is decreased
the numbers of WIMP events required in each case are increased, by
factors of $ 1.1-1.2 , \, 2-4$ and $\sim 10$ for $ S/N= 10, \, 1 $ and
$0.1$ respectively.

 For 2-d read-out with perfect angular resolution, the numbers of
 events required are a factor of roughly $2$ larger than for the same
 sense determination probability and signal to noise ratio for 3-d
 read-out with the recoil direction reconstruction uncertainty taken
 into account.  We caution that projection effects will make angular
 resolution a more significant factor for 2-d read-out than for
 3-d read-out. Estimates based on the projected length of recoil
 tracks in the planes indicate that the required number of events would
 increase by at least a further factor of 2~\cite{pap2}.

Finally we examine whether it is more important to correctly
determine the sense of the abundant low energy recoils or the rare,
but more anisotropic, high energy recoils. To do this we consider step
function sense determination probabilities: $P_{cs}(E< E_{\rm step})= 0.5
\, (1.0)$, $P_{cs}(E> E_{\rm step}) = 1.0 \, (0.5)$ for 3-d read-out and zero
background. These correspond to the sense being undetermined
(perfectly determined) for low energy recoils and perfectly determined
(undetermined) for high energy recoils. It should be noted that
neither of these possibilities, in particular the later, are
particularly physically plausible. As the energy above which the sense
can be (perfectly) determined, $E_{\rm step}$, is increased the number
of events required increases rapidly (by an order of magnitude as
$E_{\rm step}$ is increased from $20 \, {\rm keV}$ to $60 \, {\rm
  keV}$).  Conversely for the case where the sense of high energy
recoils can not be determined as $E_{\rm step}$ is decreased $N_{95}$
increases slowly initially, and then more rapidly for $E_{\rm step}<
40 \, {\rm keV}$. This indicates that determining the sense of the
common, but less anisotropic, low energy
events is most important
for minimising the number of event required to reject isotropy.

\section{Summary}

In this Brief Report we have investigated the effect of probabilistic recoil
sense determination on the number
of events required by a directional detection experiment to reject
isotropy, and detect a WIMP signal. We have considered a constant sense
determination probability and also increasing probability with
increasing recoil energy (higher energy recoils have longer tracks and
hence it is likely that it will be easier to determine their sense).
For a constant probability of
determining the sense correctly, 3-d read-out and zero background,
we find that as the probability is decreased from 1.0 to 0.75 the
number of events required to reject isotropy using the mean angle
statistic is increased by a factor of a few. As the probability is
decreased further the number of events required using this statistic
increases sharply, and in fact for $P_{100 \,{\rm keV}}<0.7$ isotropy can be
rejected more easily by discarding the sense information and using axial
statistics (this does however require an order of magnitude more events than
vectorial data with perfect sense determination).

The linearly increasing sense determination probability with $P_{cs}(E_{\rm
  th})=0.75 \, (0.50)$ and $P_{100 \, {\rm keV}}= 1.0 $ requires $\sim 3$
times (an order of magnitude) more events than the constant probability
with the same value of $P_{100 \, {\rm keV}}$. This suggests that
maximising the probability of correctly determining the sense of the
low energy, more common, but less anisotropic events is most important
for minimising the number of events required. We have confirmed this
conclusion by considering step function sense determination
probabilities.

We also considered non-zero background and 2-d read-out. As the signal
to noise ratio is decreased the numbers of WIMP events required in
each case are increased, by factors of $ \sim 1.1-1.2 , \, 2-4$ and
$10$ for $S/N= 10, \, 1 $ and $0.1$ respectively.  For 2-d read-out
with perfect angular resolution, the numbers of events required are
roughly a factor of $\sim 2$ larger than for 3-d read-out with the
recoil direction reconstruction uncertainty taken into account
(with the same sense determination probability and signal to noise ratio).

% If in two-column mode, this environment will change to single-column
% format so that long equations can be displayed. Use
%\begin{widetext}
%\end{widetext}

% figures should be put into the text as floats.
% Use the graphics or graphicx packages (distributed with LaTeX2e)
% and the \includegraphics macro defined in those packages.
% See the LaTeX Graphics Companion by Michel Goosens, Sebastian Rahtz,
% entire page. There is no need to do explicit centering.

 \begin{figure}
 \includegraphics[width=8.0cm]{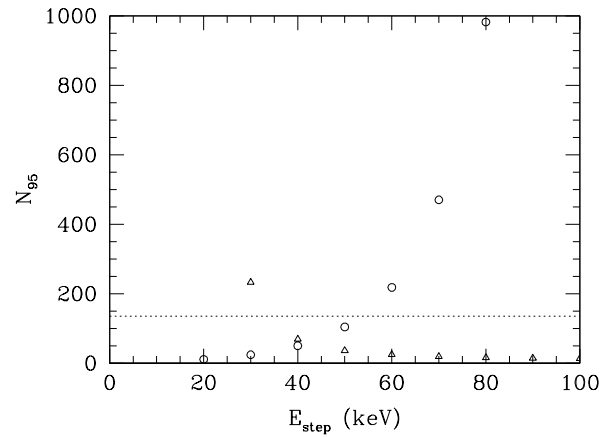}
 \caption{\label{step} $N_{95}$ for a step function probability of
   correctly determining the sense, $P_{cs}(E<E_{\rm step})=0.5 \, (1.0)$,
   $P_{cs}(E>E_{\rm step}) =1.0 \, (0.5)$ plotted using circles
   (triangles), for 3d read-out and zero background.  The dotted line
   shows the number of events required using the value of $\langle
   |\cos{\theta}| \rangle$, which does not require sense information.  }

 \end{figure}

% Surround figure environment with turnpage environment for landscape
% figure
% \begin{figure}
% \end{figure}

% tables should appear as floats within the text
% The ruledtabular enviroment adds doubled rules to table and sets a
% reasonable default table settings.
% Use the table* environment to get a full-width table in two-column
% Add \usepackage{longtable} and the longtable (or longtable*}
% environment for nicely formatted long tables. Or use the the [H]
% placement option to break a long table (with less control than 
% \begin{table}%[H] add [H] placement to break table across pages
% \begin{ruledtabular}
% Lines of table here ending with \\
% \end{tabular}
% \end{ruledtabular}
% \end{table}

% Surround table environment with turnpage environment for landscape
% table
% \begin{table}
% \begin{ruledtabular}
% \end{tabular}
% \end{ruledtabular}
% \end{table}

%\appendix
%\section{}

\begin{acknowledgments}
  AMG and BM are supported by STFC. We are grateful to Demitri Muna
  for computing assistance.
\end{acknowledgments}


\end{document}